\documentclass[aps,pra,amsmath,amssymb,twocolumn,showpacs,groupedaddress]{revtex4-1}

\usepackage{graphicx}
\usepackage{amsbsy}
\usepackage{amssymb}
\usepackage{amsfonts}
\usepackage{amsmath}
\usepackage{bm}
\usepackage{mathrsfs}
\usepackage{color}
\usepackage{hyperref}

\begin{document}

\title{Diffusive-Gutzwiller approach to the quadratically driven photonic lattice}

\author{Eduardo Mascarenhas}
\affiliation{Institute of Physics, Ecole Polytechnique F\'{e}d\'{e}rale de Lausanne (EPFL), CH-1015 Lausanne, Switzerland}

\date{\today}

\begin{abstract}
We adopt a diffusive-Gutzwiller approach to investigate a phase transition in a quadratically driven-dissipative Bose-Hubbard lattice. Diffusive trajectories may lead to lower average entanglement as compared to jump-like trajectories and have a natural tendency to approach coherent states, therefore the method can be less prone to the bias induced by the fully uncorrelated Gutzwiller ansatz. Averaging over trajectories does lead to classical correlations and this allows us to address the correlation length of such 2D lattices of open quantum systems which is the main goal of this work. Under this approximations, we find negligible correlation length in the low density phase and apparently unbounded length grow in the high density phase. Additionally, we show that the effective relaxation times associated to the times scales for synchronisation in the high density phase may also diverge suggesting the vanishing of the Lindbladian gap.\end{abstract}

\maketitle

\section{Introduction}
Driven-dissipative many-body quantum systems is a special class of nonequilibrium systems that has been through intense activity in the last decade. The possibility of genuine nonequilibrium universality classes~\cite{MarkCrit} suggests that such phenomena may not have equilibrium analogs which makes the development of nonequilibrium methodology of paramount importance~\cite{VarCui,VarL,Taiwan,iPepsRoman,BiondiCorr,CasteelsHierarchy,ClusterMF,LinkedCluster,ArrigoniImpurity,ArrigoniImpurityNE,CarusottoStochGutz,StochGutz}. On the experimental side, the realization of dissipative quantum simulators is currently not a far fetched reality in different architectures using optical or superconducting circuits \cite{FluidLight,HartmannRev,AngelakisRev}. Recently, we have witnessed experimental evidence of dissipative phase transitions in systems of ultra cold atoms \cite{Esslinger}, superconducting circuits \cite{Circuits}, and semiconductors \cite{Semicond}.

Lattices of photonic nonlinear resonators have been studied in the last decade even since the low temperature analogy to condensed matter was established, suggesting the possibility of simulating the Mott-Superfluid phase transition~\cite{MottSup}. However, the equilibrium Mott-Superfluid transition is not resilient to the typical highly dissipative nature of such systems. The study of this phenomenology has progressed along two main lines. The first is the nonequilibrium synthesis and stabilization of strongly correlated phases such as Mott~\cite{VFriends} and solid phases~\cite{Solid} in analogy to the
ground state physics. The second is the identification and characterization of genuine nonequilibrium phases and transitions such as the emergence of the gas-liquid bistability for the linearly driven case~\cite{BiondiGas,Vicentini}, spontaneous symmetry breaking for the quadratically driven case~\cite{Vincenzo} and laser like transitions driven by interactions~\cite{Laser}. In this context, the quadratically driven single Kerr resonator has been studied, both theoretically~\cite{Quad1,Quad2,Quad3,Quad4,Quad5,Quad6,Quad7,Quad8} and experimentally~\cite{QuadEx}. 

Theoretically addressing two-dimensional driven dissipative lattices is, in general, a difficult task. The local dimension of the vector space is usually hight as compared to small spin systems. For example, keeping ten Fock states results in a density matrix of local dimension 100, making it hard to model such high occupation systems with tensor-network techniques~\cite{VarCui,VarL,iPepsRoman,FabianBose}. Such techniques may also generate negative density matrices requiring extra resources to ensure positivity~\cite{Positive}. It is also not clear to what degree quantum correlation are relevant in such bulk-driven-dissipative lattices. An alternative, that captures only classical correlations and scales linearly with system size, is the stochastic Gutzwiller approach~\cite{CarusottoStochGutz,StochGutz}. The method has been recently applied to a spin model~\cite{StochGutz} in which the trajectories are driven by Poisson noise, thus being of jump-like behaviour. The only limitations of the method are (i) that the results can be biased by the fact that quantum correlations are neglected in each trajectory and (ii) it can happen that the jump trajectories do not spontaneously create a coherent fraction leading to an unphysical behaviour in which the sites do not ``talk" to each other under the mean-field-like coupling. 
However, It has been known that, markovian master equations may be unraveled in infinitely many ways and that diffusive local unravelings may lead to lower average entanglement as compared to jump unravelings~\cite{LocalDiff}. In the specific case addressed in~\cite{LocalDiff} local diffusion leads to the minimum average entanglement obtained with local unravelings. Furthermore, in the specific case of the quadratically driven resonator addressed here, local diffusion tends to generate trajectories with significant coherent fraction as shown in~\cite{Quad2}. Such features make the diffusive version of the stochastic-Gutzwiller approach possibly (i) less prone to biases due to the lower entanglement of diffusive trajectories and (ii) less prone to unphysical behaviour of the jump version since the noise in the diffusive trajectories allows for communication between sites through a fluctuating coherent fraction. It should be noted, however, that we do not expect the method to be the ultimate tool for such 2D lattices but rather a useful alternative.

In this work we review the theory behind the method applying the time depend variational principle~\cite{TDVP} to the stochastic Schrodinger equation while confining each trajectory to the manifold of uncorrelated states (Gutzwiller ansatz). This short overview of the theory putting the method on firmer grounds is done in Section II. Subsequently, we present the application of the method to the quadratically driven photonic lattice in Section III in which we presents the results of the numerical analysis addressing issues of dimensionality, correlation lengths and and relaxation times as we cross the transition. We find that both the correlation length and the relaxation rates present signatures of the phase transition. In Section IV we present our conclusions.

\section{Theory}

This work is devoted to open quantum systems whose density matrix obey a Markovian master equation of Lindblad form such as~\cite{QNoise}
\begin{equation}\frac{d\rho}{dt}=-i[H,\rho]-\frac{1}{2}\sum_i \left[  K_i^{\dagger}K_i\rho+\rho K_i^{\dagger}K_i -2K_i\rho K^{\dagger}_i\right],\end{equation}
in which $H$ is the system hamiltonian and $K_i$ are so called jump operators through which the system couples to the environment.
Such master equation may be unraveled with a stochastic Schrodinger equation typically found in homodyne measurements~\cite{QMAC}
\begin{eqnarray}d|\tilde{\Psi}\rangle&=&-iH_{\mathrm{eff}}[\mathbf{Q}]dt|\tilde{\Psi}\rangle\nonumber\\
&=&\left[-iHdt  -\frac{1}{2}\sum_i K^{\dagger}_iK_idt+\sum_iK_idQ_i\right]|\tilde{\Psi},\rangle\label{Diff}\end{eqnarray}
with $dQ_i=\langle K^{\dagger}_i+ K_i\rangle dt + dW_i$ being the homodyne currents and $dW_i$ being independent Wiener processes. The above equation is written in a format appropriate for efficient numerical integration and does not preserve norm such that $|\tilde{\Psi}\rangle$ is an unormalized version of $|\Psi\rangle$. Typically the state is renormalized after each time step.

The time dependent variational principle is based on minimization of the functional 
\begin{equation} f=\left\langle \Psi\left| \frac{\partial}{\partial t}+iH_{\mathrm{eff}}[\mathbf{Q}]\right|{\Psi}\right\rangle\end{equation}
such that equation~(\ref{Diff}) is recovered from setting the functional derivative to zero $\frac{\delta f}{\delta \langle {\Psi}|}=0$. The Gutzwiller ansatz in this context consists of restricting the dynamics of each trajectory to the manyfold of uncorrelated states $|\Psi\rangle=\bigotimes_s|\psi_s\rangle$ such that $|\psi_s\rangle$ is a pure state of site $s$. Let also assume that equation~(\ref{Diff}) only has two body couplings such that  
$H_{\mathrm{eff}}[\mathbf{Q}]=\sum_{s'\le s}H^{(s)}_{s'}[\mathbf{Q}]$ (where the ordered sum excludes double counting).
The functional derivative $\frac{\delta f}{\delta \langle {\psi}_s|}=0$ leads to the following set of coupled equations of motion 
\begin{equation} d|\tilde{\psi}_s\rangle=-i\left[H^{(s)}_{s}|\tilde{\psi_s}\rangle
+\sum_{s'\ne s}\langle \psi_{s'}| H^{(s)}_{s'}|\psi_{s'}\rangle\right]|\tilde{\psi_s}\rangle dt,\end{equation}
such that $\langle \psi_{s'}| H^{(s)}_{s'}|\psi_{s'}\rangle$ should be immediately identified as the mean-field-like effective hamiltonian that couples site $s$ to site $s'$. Finally, after averaging over trajectories, or equivalently, averaging over the Gaussian noise we recover a classically correlated state
\begin{equation}\rho_{CC}=\int d\mu(\mathbf{Q})|\Psi[\mathbf{Q}]\rangle\langle\Psi[\mathbf{Q}]|=\overline{|\Psi[\mathbf{Q}]\rangle\langle\Psi[\mathbf{Q}]|}\end{equation}
 that is as an approximation of the full quantum state $\rho=\rho_{CC} +\rho_{QC}$, 
such that $\rho_{QC}$ represents the quantum correlated component. Now, the method is only expected to be accurate in lattices of high connectivity and in case $\rho_{QC}$ is negligible. Testing this hypothesis remains an open challenge and the classically correlated ansatz for $\rho$ remains ultimately as an uncontrolled approximation.

\section{The quadratically driven Bose-Hubbard model}

The effective Hamiltonian components of the quadratically driven Bose-Hubbard model are given by 
\begin{eqnarray}
H^{(s)}_{s'} &=&\Delta a^{\dagger}_sa_s+\frac{U}{2}a^{\dagger}_sa^{\dagger}_sa_sa_s+\frac{G}{2}(a^{\dagger 2}_s+a^{2}_s)\nonumber \\
&-&\frac{i}{2}\sum_i^2 K^{(s)\dagger}_iK^{(s)}_i+i\sum_i^2K^{(s)}_i\frac{dQ^{(s)}_i}{dt}\quad \mathrm{if} \  s=s'\nonumber\\
&=&-J_{\langle s,s'\rangle}\left(a_sa^{\dagger}_{s'}+a^{\dagger}_sa_{s'}\right) \quad \mathrm{if} \ s\ne s'
\label{H}
\end{eqnarray}
where $a_s$ is the bosonic annihilation operator for the $s$-th site, $J$ is the hopping strength,
In this expression, $U$ is the strength of the Kerr nonlinearity, $G$ is the amplitude of the two-photon driving. In the rotating frame, $\Delta=\omega_c-\omega_2/2$, where $\omega_2$ is the frequency of the two-photon driving and $\omega_c$ is the resonator frequency. The jump operators at each site are $K^{(s)}_1=\sqrt{\gamma_1}a_s$ and $K^{(s)}_1=\sqrt{\gamma_2}a^{2}_s$, with $\gamma_1$ and $\gamma_2$ being the single and two photon dissipation rates, respectively.

\begin{figure}
{\includegraphics[width = 2.5in]{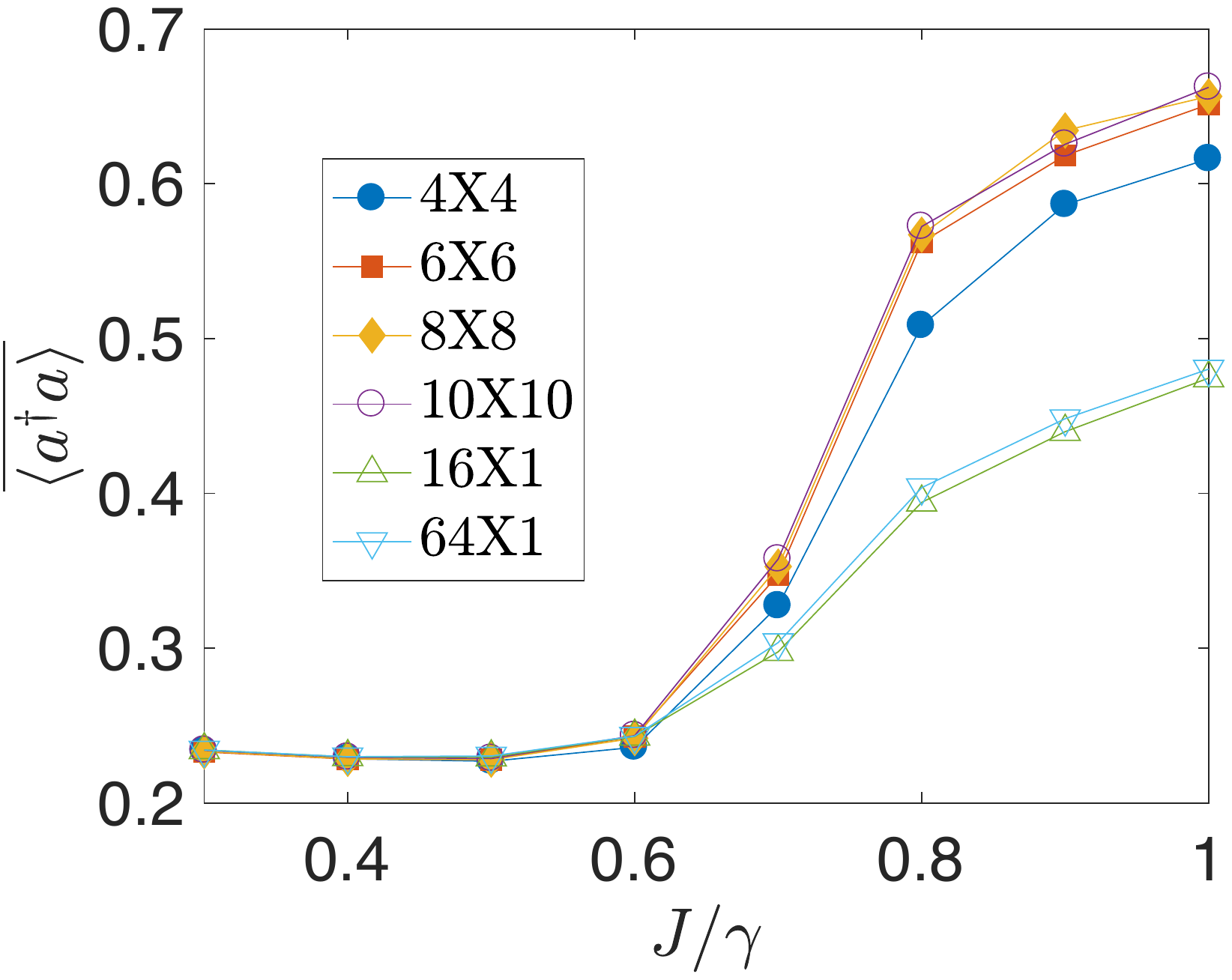}}
\caption{Spatially averaged photon number generated with the DG method as a function of hopping strength for both 1D and 2D geometries. Parameters are chosen as $\gamma_1=\gamma_2=\gamma$, $G=4\gamma$, $U=10\gamma$ and $\Delta=J$.}
\label{photon}
\end{figure}

\begin{figure}
{\includegraphics[width = 3in]{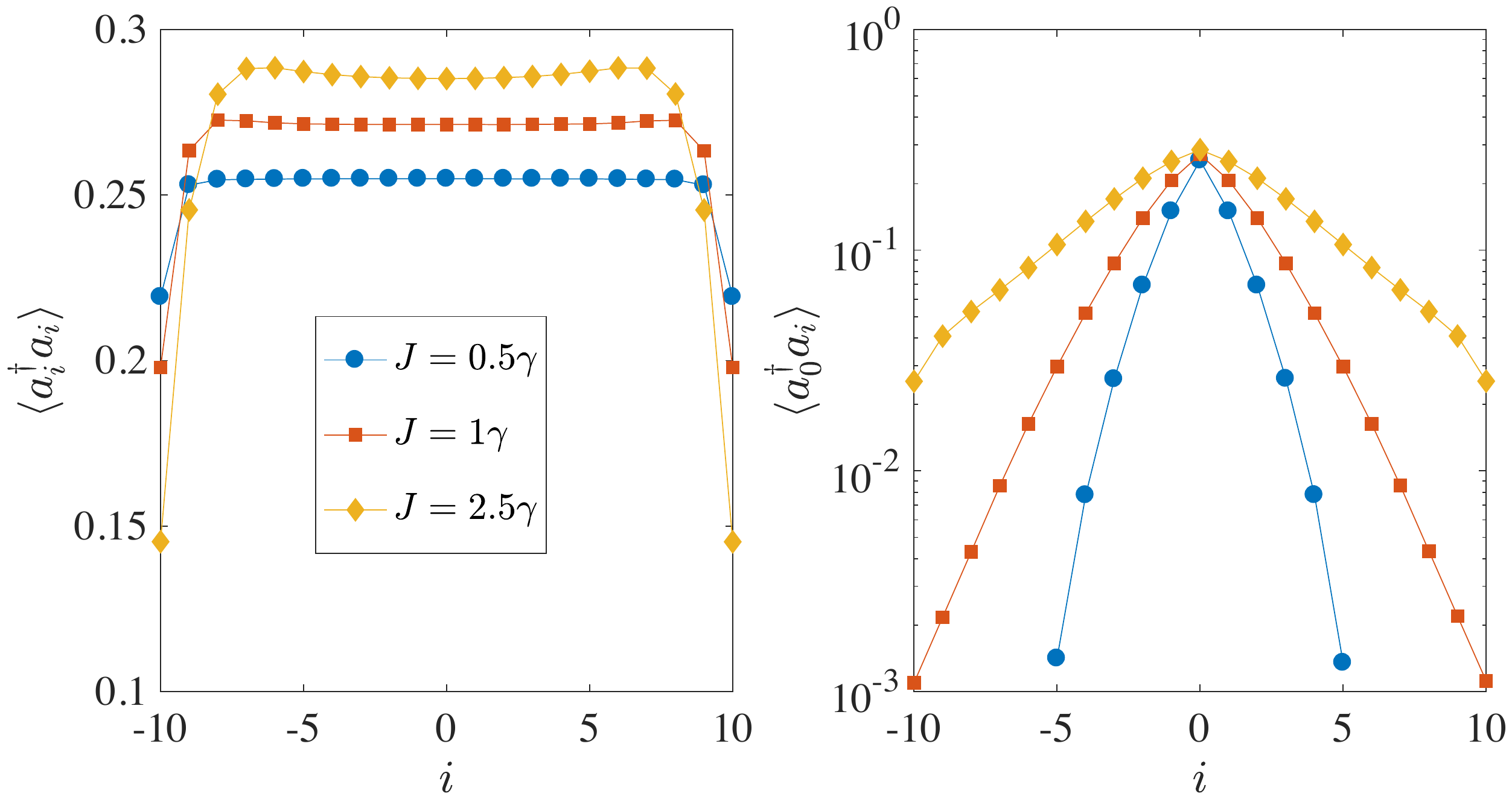}}
\caption{ (Left) Average photon number and (Right) first order correlations generated with an MPS approach. Parameters are chosen as $\gamma_1=\gamma_2=\gamma$, $G=4\gamma$, $U=10\gamma$ and $\Delta=J$.}
\label{MPS}
\end{figure}

In figure~\ref{photon} we show the asymptotic photon number averaged in real space simulated with the Diffusive-Gutzwiller (DG) method in reasonable agreement with the mean-field uncorrelated method in~\cite{Vincenzo} with the estimated transition point being around $J/\gamma\approx 0.6$. A direct comparison between 1D and 2D (square lattice) results suggests that dimensionality plays an import role in this system. However, it must be stressed that the DG results are unreliable in one dimension as is typically the case with mean-field-like methods. A direct comparison with a matrix-product-state simulation in figure~\ref{MPS} shows that the DG method over estimates the photon number in 1D chains in the higher density (or symmetry broken) phase, also suggesting that short range quantum correlations may be more relevant in the 1D geometry.

\begin{figure}
{\includegraphics[width = 3.4in]{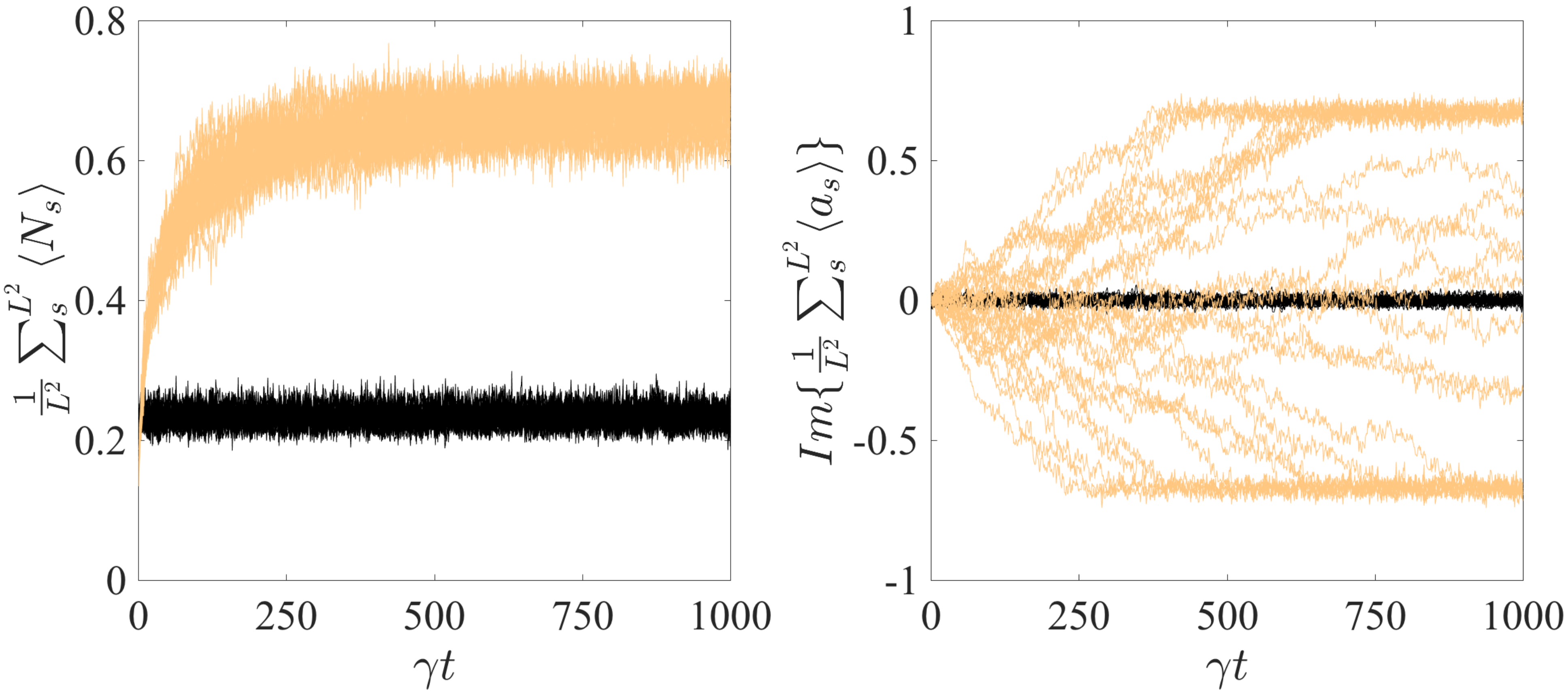}}
\caption{ DG simulations for a 2D system $L$x$L$ with $L=11$. (Left) The average total photon number and (right) the average total coherent component. Dark and bright curves correspond to $J/\gamma=0.3$ and $J/\gamma=1$, respectively. Parameters are chosen as $\gamma_1=\gamma_2=\gamma$, $G=4\gamma$, $U=10\gamma$ and $\Delta=J$.}
\label{TrajJp3J1}
\end{figure}

Let us now take a closer look at single realizations of the noise process. In figure~\ref{TrajJp3J1} we show several trajectories for one point in the low density phase ($J=0.3\gamma$) and one point in the high density phase ($J=1\gamma$). Here we define the ``macroscopic" quantities that average local quantities over the 2D spacial degrees of freedom. The average total photon number $N=\frac{1}{L^2}\sum_s^{L^2}\langle N_s\rangle$ and average total coherent component $\alpha=\frac{1}{L^2}\sum_s^{L^2}\langle a_s\rangle$. The photon number has an obvious local interpretation, however the coherent component actually captures a global feature regarding the phase synchronization of the oscillators. If the oscillators are out of phase, which is the case in the low density phase, $|\alpha|=0$ such that their phases randomly average to zero, while in the high density phase the oscillators synchronize presenting a global phase locking yielding $|\alpha|>0.$ In figure~\ref{TrajJp3J1} we show the slow buildup of the coherent component for several trajectories allowed by the synchronization in the high density phase and the absence thereof in the low density phase. The simulations are started in the vacuum state which has a pathological  behaviour in the quantum jump approach. Note that none of this features would be observable with the quantum-jump approach since the jump trajectories never spontaneously generate a local coherent component ($\langle a_s\rangle$) under the quadratic driving as deeply discussed in~\cite{Vincenzo,Quad2}, which forbids communication between site under the Gutzwiller approximation. In the diffusive approach, the terms $KdQ$ in equation~(\ref{H}) add fluctuations to the local coherent component allowing for communication between the sites and eventually the synchronization. It should also be noted that in figure~\ref{TrajJp3J1} we observe the symmetric asymptotic solutions as predicted in~\cite{Vincenzo}, however while averaging over trajectories we recover a unique steady state with zero coherent component ($\mathrm{tr}\{ \rho a_s \}=0, \forall s$) since the negative and positive trajectories cancel each other. The synchronisation does manifest itself in the spacial correlation functions which we will address.

\begin{figure}
{\includegraphics[width = 3.4in]{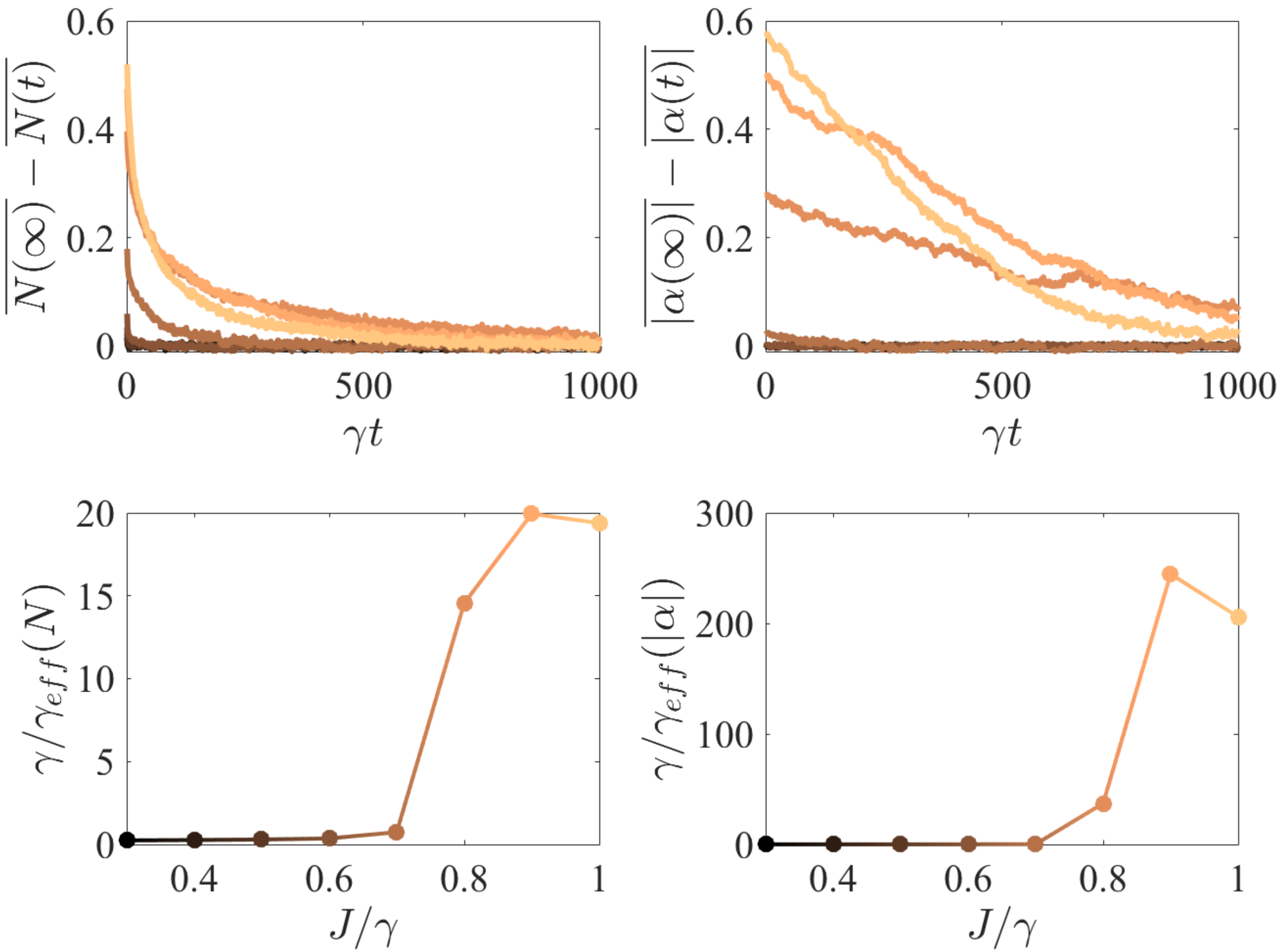}}
\caption{ DG simulations for a 2D system $L$x$L$ with $L=11$. (Upper-left) The relaxation dynamics of the average total photon number and (upper-right) the average total coherent component. The effective relaxation times for the (lower-left) photon number and (lower-right) coherent component. Dark curves correspond to lower values of $J/\gamma$ and bright curves correspond higher values of $J/\gamma$. Parameters are chosen as $\gamma_1=\gamma_2=\gamma$, $G=4\gamma$, $U=10\gamma$ and $\Delta=J$.}
\label{Relax}
\end{figure}

Directly probing the gap of the Lindbladian is a desirable task since it dictates the effective relaxation rate that emerges from the many-body dynamics. However, it is not directly accessible with the DG method. Effective relaxation rates my be inferred from the relaxation of observables such as $O(\infty)-O(t)\approx e^{-\gamma_{\mathrm{eff}}(O)t}$, however the choice of observable may dramatically influence the results such that different observables $O$ may have different relaxation rates $\gamma_{\mathrm{eff}}(O)$. Here we study the relaxation of both the photon number and the coherent component. In figure~\ref{Relax} we show the relaxation dynamics and effective relaxation rate for both quantities and we find that the rates are very small in the high density phase indicating the vanishing of the Lindblad gap. Furthermore, the relaxation time of the coherent component can be one order of magnitude larger then the photon number relaxation time. This shows how global quantities that are affected by the synchronisation take longer to reach their asymptotics and that global quantities should be used in order to obtain a more accurate estimation of the Lindblad gap.

\begin{figure}
{\includegraphics[width = 3.3in]{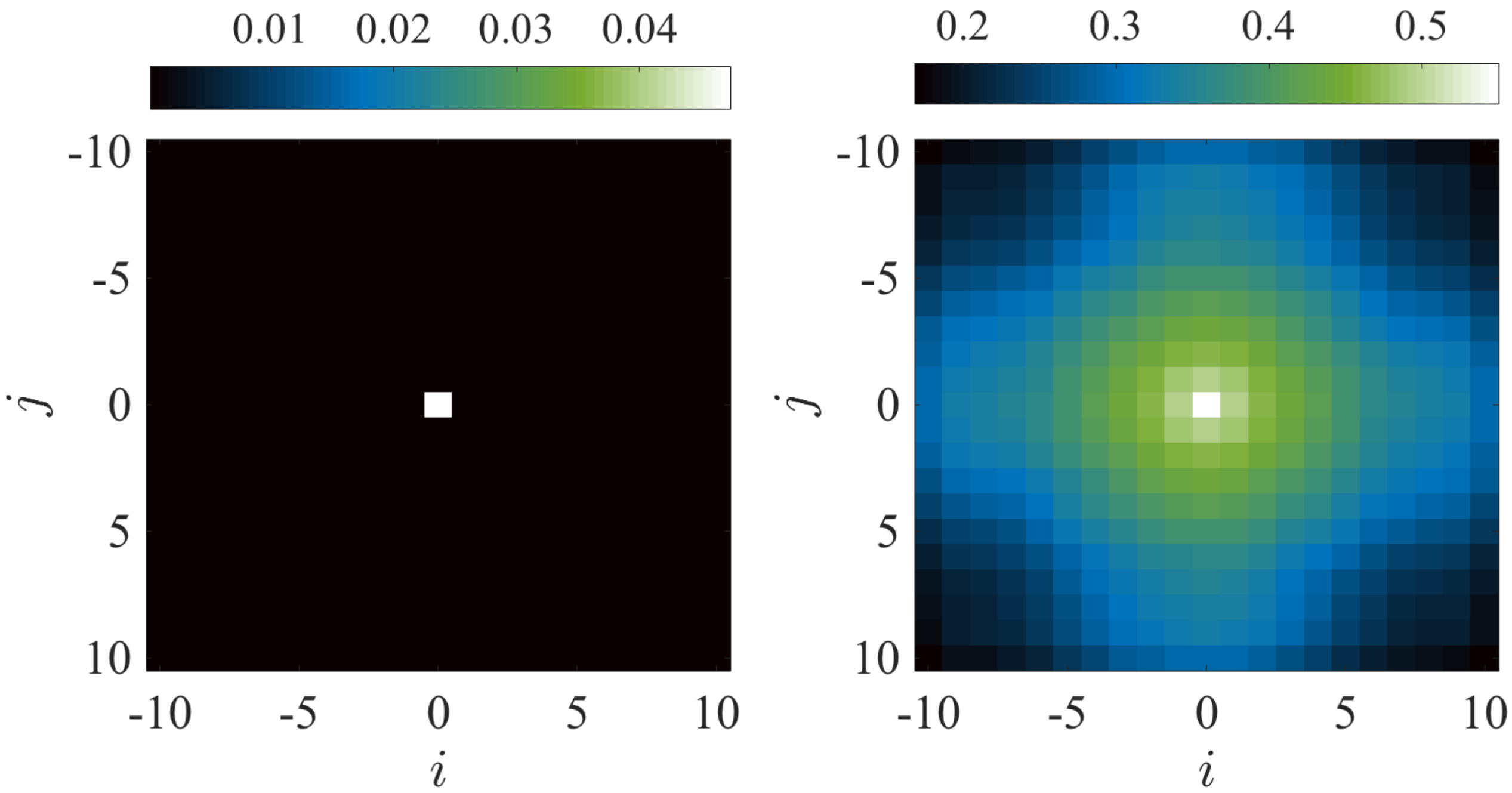}}
\caption{ First order correlation function of the central with the rest of the lattice for a 2D system $L$x$L$ with $L=21$ for (Left) $J/\gamma=0.3$ and (Right) $J/\gamma=1$. Parameters are chosen as $\gamma_1=\gamma_2=\gamma$, $G=4\gamma$, $U=10\gamma$ and $\Delta=J$.}
\label{g1U10G4L21}
\end{figure}

Under the DG approximation the first order correlation assumes a classical nature $\langle a^{\dagger}_sa_{s'}\rangle\approx \overline{\langle a^{\dagger}_s\rangle \langle a_{s'}\rangle}$. We are specifically interested in the correlation between the site in the middle of the 2D lattice and all other sites 
\begin{equation}G_{i,j}=\left|\overline{\langle a^{\dagger}_{0,0}\rangle \langle a_{i,j}\rangle}\right|.\end{equation} In figure~\ref{g1U10G4L21} we show $G$ both in the low and high density phases. in the low density phase we observe essentially only self-correlation while in the higher density phase the correlation expands over the whole lattice. The difference in protonic density between the two phase is so very drastic in the parameter regimes studies here. Thus, it is remarkable how the synchronization of the oscillators in the high density phase overcomes the local dissipation mechanisms and generates long range coherence similar to a super-fluid state.

\begin{figure}
{\includegraphics[width = 3.5in]{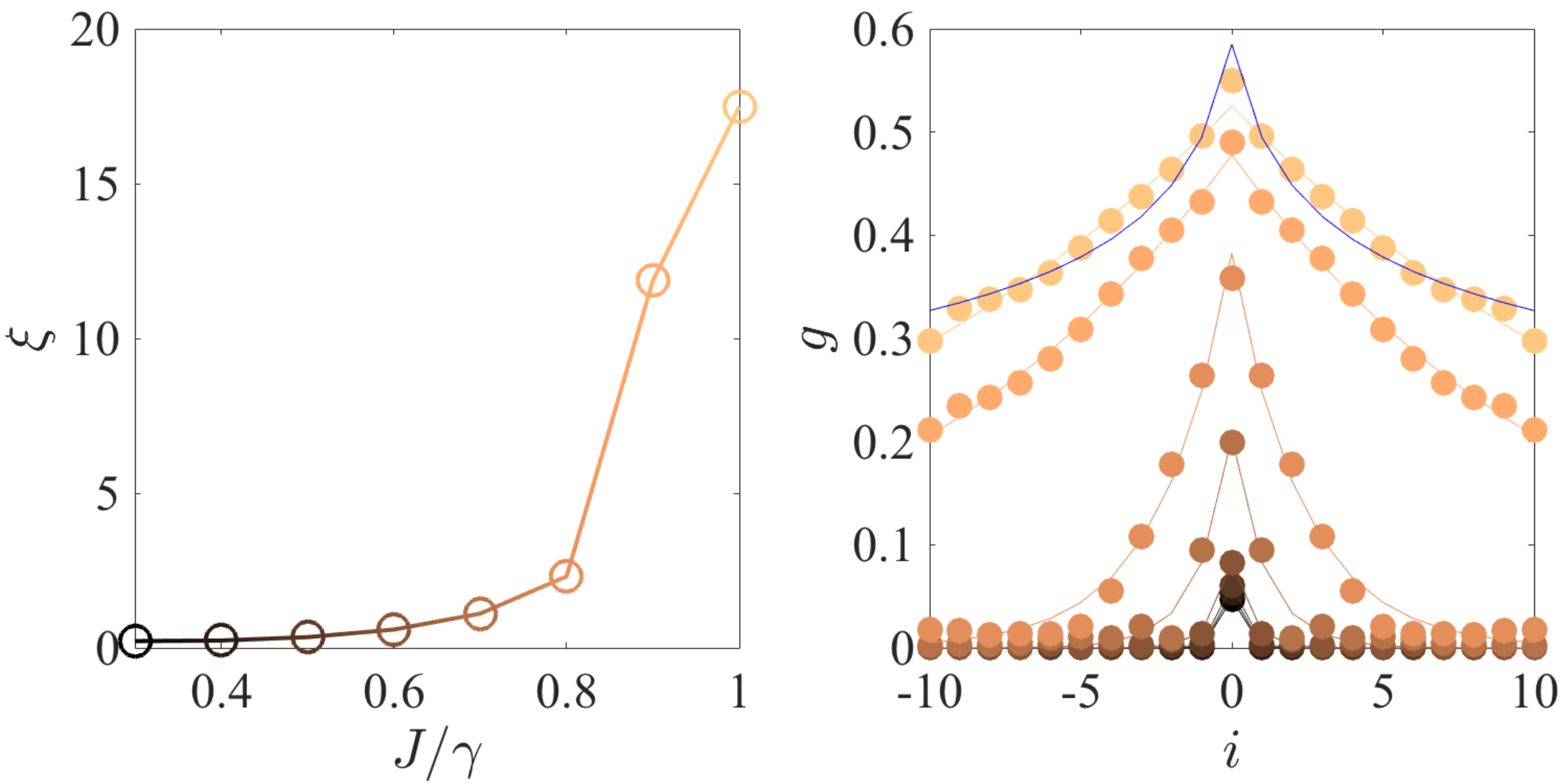}}
\caption{ DG simulation for a system $L$x$L$ with $L=21$.(Left) Exponential correlation length and (Right) the corresponding $g_{0,i}$ correlation functions. The dots in the right panel are the DG data while the same color line are the exponential fits $ce^{-|i|/\xi}$ and the blue line is a power law fit $c|i|^{-\xi}$ to the correlation function at $J/\gamma=1$. Parameters are chosen as $\gamma_1=\gamma_2=\gamma$, $G=4\gamma$, $U=10\gamma$ and $\Delta=J$. Bright color correspond to higher values of $J/\gamma$ while dark colors to lower values.}
\label{Length11g1U10G4}
\end{figure}

In order to allow for direct comparisons with the 1D case let us define a 1D projection of the correlation function as
\begin{equation} g_{0,i}=\frac{G_{0,i}+G_{i,0}}{2}.\end{equation}
We may also define the effective correlation length assuming $g_{0,i}\propto e^{-|i|/\xi}$.
In figure~\ref{Length11g1U10G4} we show the correlation length as a function of the hopping strength and $g_{0,i}$ for several values of $J/\gamma$ as we cross the transition. Negligible correlation length is found in the low density phase while in the high density phase the correlation length becomes approximately the system size $\xi\approx L$ at least for the maximal system size we have been able to simulate $L=21$. From our results we see no indication that the correlation assumes a power law shape (as is the case for the equilibrium super-fluid phase). We figure~\ref{Length11g1U10G4} we fit all the correlation functions with an exponential and find the fits to be accurate. For the sake of comparison we also show a power law fit to the correlation function at $J/\gamma$. The exponential decay of the correlation is also present in 1D as shown in figure~\ref{MPS} with an MPS approach. This results also raise the question to weather or not the transition predicted in~\cite{Vincenzo} persists in presence of correlations or if the universality class is altered. These issues are, however, difficult to address requiring long simulation campaigns and possibly improved ansatz that take quantum correlations into account on short length scales.

\section{Conclusions}

We have carried out a study of the quadratically driven photonic lattice incorporating classical correlations with the diffusive-Gutzwiller approach. We have observed the growth of both relaxation times and correlation lengths in the high density regime in remarkable contrast to the low density regime. Determining precisely the transition points and universality classes might require taking quantum correlations into account because these are relevant at short length scales and influence the the region at which correlations starts to spread and give rise to the emergent transition point.
Since the method does not capture quantum correlations, it could be of considerable relevance to account for quantum correlations at least on short length scales. This could be pursued with different techniques, but also still under the Gutzwiller umbrella considering complementary cluster states. We will pursue this with the variational principle outlined in this work.

\begin{acknowledgements}
Vincenzo Savona is acknowledged for several insightful discussions and precious suggestions while supporting this project.
\end{acknowledgements}

\end{document}